\documentclass[12pt]{article}
\usepackage{amssymb,amsmath,epsfig,cite}
\allowdisplaybreaks 
\allowdisplaybreaks 
\begin{document}
\title{\textbf{Formation of Cylindrical Gravastars in Modified Gravity}}
\author{Z. Yousaf$^1$ \thanks{zeeshan.math@pu.edu.pk}, Kazuharu Bamba$^2$
\thanks{bamba@sss.fukushima-u.ac.jp} and M. Z.
Bhatti$^1$ \thanks{mzaeem.math@pu.edu.pk}\\
$^1$ Department of Mathematics, University of the Punjab,\\
Quaid-i-Azam Campus, Lahore-54590, Pakistan\\
$^2$ Division of Human Support System,\\ Faculty of Symbiotic
Systems Science,\\ Fukushima University, Fukushima 960-1296,
Japan}

\date{}

\maketitle
\begin{abstract}
In this paper, we analyze a few physical characteristics of
gravastar with cylindrical geometry in $f(R,T)$ theory, where $R$ is
the Ricci scalar and $T$ is the trace of energy-momentum tensor. The
gravastar is generally considered to be a substitute of a black hole
with three different regions. In the present work, we examine the
formulation of gravastar-like cylindrical structures in $f(R,T)$
theory. By using Darmois and Israel matching conditions, we
formulate a mass function of a thin shell. We calculate the
different physical characteristics of gravastar, in particular,
entropy within the thin shell, proper length of the intermediate
thin shell as well as energy of the shell.
\end{abstract}
\section{Introduction}

Some current experimental and observational results of cosmos
\cite{ya1} asserted that our universe is expanding with an
acceleration. Moreover, it has been analyzed that our universe is
comprised of 5\% baryonic matter along with 27\% and 68\% as dark
matter (DM) and dark energy (DE), respectively \cite{Ade:2014xna,
Ade:2015tva, Array:2015xqh}. Apart from general relativity (GR),
there are many popular ways to comprehend evolutionary phases of our
cosmos in modified theories, the theories that could be formulated
from the modification of GR action function. Nojiri and Odintsov
\cite{ya3} indicated how these gravity theories provide help in the
study of cosmic structure formation. Recent attractive modified
theories include $f(R)$ with $R$ being the Ricci scalar \cite{fR1},
$f(\mathcal{T})$ (here, $\mathcal{T}$ represents the torsion scalar)
\cite{fT1}, $f(R,\Box R, T)$ (where, $T$ being the trace of matter
tensor while $\Box$ indicates the de Alembert's operator)
\cite{box1}, $f(G)$ with $G$ as the Gauss-Bonnet term \cite{fG} and
$f(G,T)$ \cite{fGT}, are the most popular gravity theories as
alternative to GR (please see \cite{R-NO-CF-CD} for details). Harko
\textit{et al.} \cite{harko1} presented $f(R,T)$ gravity by invoking
corrections of $T$ in $f(R)$ gravity. They took $T$ in their
calculations in order to study quantum effects. Houndjo
\cite{hound1} used this theory and presented some important
cosmological models that are applicable at a certain cosmic era. The
concept of Gravastars was introduced through the theoretical
modeling of Mazur and Mottola \cite{1,2} by generalizing the basic
concept of condensation provided by Bose-Einstein (BEC) for which the gravastar is governed by following regions.\\
1.  Interior cylindrical region:~~~~~~~~~~~~$p=-\rho$,\\
2.  Transitional layer: ~~~~~~~~~~~~~~~$p=+\rho$,\\
3.  Exterior cylindrical region:~~~~~~~~~~$p=\rho=0$.\\
In general, some other equations of state are described by Visser
and Wiltshire \cite{visser1}. The gravastars's shell is treated here
to be very thin ranging $r\in(r_1,r_2)$, here $r_1$ and $r_2$ are
equivalent to $D$ and $D+\epsilon$ as the interior and exterior
radial values. A lot of work is available in literature about the
gravastar and their physical characteristic \cite{new1}.

Ghosh et al. \cite{4} found a class of solutions under two
considerations for a $d$-dimensional metric with Einstein-Maxwell
corrections. They discussed few captivating outcomes from the
Reissner-Nordstr\"{o}m model. Horvat et al. \cite{5} proposed a
solution of gravastar under the influence of charged field. For the
existence of such structure, they studied some corresponding
astrophysical results. They calculated charged equations of motion,
energy conditions and equation of state by making use of de-Sitter
geometry as an interior geometry and Reissner-Nordstr\"{o}m (RN) as
outer metric.

DeBenedictis et al. \cite{8} investigated the existence of
gravastars with the help of qualitative analysis and presented some
viability bounds through energy conditions and a particular EoS.
Carter \cite{9} also discussed the stability of such bodies with the
help of few conditions in order to model thin shell. The obtained
solutions were found after taking a de-Sitter geometry as an inner
region and Reissner-Nordstr\"{o}m geometry as an outer metric.
Chirenti and Rezzolla \cite{10} described the criteria by which one
can perform differentiation of gravastar from a black hole. They
calculated quasinormal constraints in order to present a generic
gravastar models. Cecilia et al. \cite{12} described some physical
properties of for the fast rotating celestial bodies and claimed
that there could be no horizon for the possible formation of
gravastars.

Recently, Bhatti and his collaborators \cite{a} found different
significantly crucial results related to the collapse and
instability of compact objects. Yousaf \cite{sg1} calculated the
wide range of parametric values under which the obtained cylindrical
gravastar solutions satisfy the stability conditions. He presented
various qualitative aspects of EoS with the help of these
parameters. His work is then extended for
$f(R,T,R_{\mu\nu}T^{\mu\nu})$ \cite{sg2} and $f(R,G)$ \cite{sg3}
theories of gravity with spherical spacetimes. Recently, the origin as well as the
existence of gravastars have been reviewed by Ray et al. \cite{rayrev1}.

This paper extended the work of \cite{26} and \cite{u1} for
cylindrically symmetric metric in $f(R,T)$ theory. The work of this
paper is arranged in following sections. In section \textbf{2}, we
formulate $f(R,T)$ field equation and few important corresponding
equations. In section \textbf{3} and \textbf{4}, we calculate the
hydrostatic equation and gravitational mass with interior geometry.
In section \textbf{5}, we shall elaborate on the relation between
pressure and radius. In section \textbf{6}, we calculate the mass of
the shell by matching inner and outer regions of spacetime. Section
\textbf{7} describes a few important characteristics for the
existence of gravastars. In section \textbf{7}, we will discuss the
various physical characteristic of gravastar.

\section{Formulation of $f(R,T)$ Gravity}

This section describe basic framework of $f(R,T)$ theory as well as
their conservation equation. The action of $f(R,T)$ \cite{14} is
characterized with a generic function $f(R,T)$ as follows
\begin{eqnarray}\label{1}
\textbf{S}=\frac{1}{16\pi}\int\left(16\pi\textit{\L}_m\sqrt{-g}+{\sqrt{-g}f(R,\textit{T})}\right)~d^4x,
\end{eqnarray}
and its trace by $T=g^{\xi\eta}T_{\xi\eta}$. The matter tensor can
be evaluated as \cite{15}
\begin{eqnarray}\label{2}
T_{\xi\eta}=\frac{-2}{\sqrt{-g}}\frac{\delta(\sqrt{-g}\textit{\L}_m)}{\delta
g^{\xi\eta}},
\end{eqnarray}
where $\textit{\L}_m$ indicates the matter Lagrangian and depends
upon the metric $g_{\xi\eta}$ as
\begin{eqnarray}\label{3}
T_{\xi\eta}=g_{\xi\eta}\textit{\L}_m-2\frac{\partial\textit{\L}_m}{\delta
g^{\xi\eta}}.
\end{eqnarray}
The variation on the action of the gravity corresponding to the
metric $g_{\xi\eta}$ yields
\begin{eqnarray}\label{4}
\delta S=\frac{1}{16\pi}\int\left[f_R\delta R+f_T\frac{\delta
T}{\delta g^{\xi\eta}}\delta
g^{\xi\eta}-\frac{1}{2}g_{\xi\eta}f\delta
g^{\xi\eta}+16\pi\frac{1}{\sqrt{-g}}\frac{\delta(\sqrt{-g}\textit{\L}_m)}{\delta
g^{\xi\eta}} \right]\sqrt{-g}~d^4x.
\end{eqnarray}
The change in the Ricci scalar $R$ can be given as follows
\begin{eqnarray}\label{5}
\delta R=\delta(g^{\xi\eta}R_{\xi\eta})=R_{\xi\eta}\delta
g^{\xi\eta}+g^{\xi\eta}(\nabla_\omega\delta\Gamma^\omega_{\xi\eta}-\nabla_\eta\delta\Gamma^\omega_{\xi\omega}),
\end{eqnarray}
where $\nabla_\omega$ represents covariant differential operator. The variation of Christoffel
symbol provides
\begin{eqnarray}\label{6}
\delta\Gamma^\omega_{\xi\eta}=\frac{1}{2}g^{\omega\alpha}(\nabla_\xi
\delta g_{\eta\alpha}+\nabla_\eta \delta g_{\alpha\xi}-\nabla_\alpha
\delta g_{\xi\eta}).
\end{eqnarray}
Using the above equations in Eq.\eqref{4}, we get field equation of
$f(R,T)$ theory as
\begin{eqnarray}\label{7}
f_{R}(R_{\xi\eta}-\nabla_\xi\nabla_\eta+g_{\xi\eta}\Box)-\frac{1}{2}f
g_{\xi\eta} +f_{T}(T_{\xi\eta}+\Theta_{\xi\eta})=8\pi T_{\xi\eta},
\end{eqnarray}
where $f$=$f(R,\textit{T}),~f_{R}$ and $f_{T}$ are the partial
differentiations of arbitrary function corresponding to $R$ and $T$
respectively, $\Box$=$\nabla^2$=$\nabla_\eta\nabla^\eta$ and
\begin{eqnarray}\label{8}
\Theta_{\xi\eta}=\frac{g^{\alpha\beta}\partial
T_{\alpha\beta}}{\partial g_{\xi\eta}}.
\end{eqnarray}
The covariant derivative of Eq.\eqref{7} is \cite{16}
\begin{eqnarray}\label{9}
\nabla^\xi
T_{\xi\eta}=\frac{f_{T}(R,\textit{T})}{8\pi-f_{T}(R,\textit{T})}[(T_{\xi\eta}+\Theta_{\xi\eta})\nabla^\xi\ln{f_T(R,\textit{T})}+\nabla^\xi(\Theta_{\xi\eta}
-\frac{1}{2}g_{\xi\eta}T)].
\end{eqnarray}
We want to consider the isotropic cylindrically symmetric spacetime
in order to model gravastars in $f(R,T)$ theory. Thus, we take
matter tensor for the isotropic fluid as
\begin{eqnarray}\label{10}
T_{\xi\eta}=(p+\rho)U_{\xi}U_{\eta}- pg_{\xi\eta},
\end{eqnarray}
where $U_{\xi}$ is the four velocity component satisfying
$U_{\xi}U^{\xi}=1$, $\rho$ indicates matter density while $p$ shows
the isotropic pressure of the fluid. In this environment we took the
specific choices of few quantities as $\textit{\L}_m=-p$ and
$\Theta_{\xi\eta}$=$-(2T_{\xi\eta}+p g_{\xi\eta})$ along with
$f(R,T)$=$R\left(1+\frac{2\chi T}{R}\right)$ with constant term
$\chi$. By substituting these, Eq.\eqref{7} becomes
\begin{eqnarray}\label{11}
G_{\xi\eta}=8\pi T_{\xi\eta}+\chi[2T_{\xi\eta}+g_{\xi\eta}(T+2p)],
\end{eqnarray}
where $G_{\xi\eta}$ represents the Einstein tensor and $T$ the trace
of matter tensor. One can write Eq.\eqref{9} as
\begin{eqnarray}\label{12}
\nabla^\xi
T_{\xi\eta}=\frac{-\chi}{2(4\pi+\chi)}\left[2\nabla^\xi(pg_{\xi\eta})+g_{\xi\eta}\nabla^\xi
T\right].
\end{eqnarray}
Using $\chi=0$ in Eq.\eqref{11}, one can easily get the field
equation of $GR$.

\section{Formulation of Field Equations with Cylindrically Symmetric Spacetime}

We take static cylindrical metric
\begin{eqnarray}\label{13}
ds^2=-dt^2H(r)+dr^2K(r)+(d\phi^2+\alpha^2dz^2)r^2,
\end{eqnarray}
where $H(r)=\sqrt{\alpha^2r^2-\frac{4M}{\alpha r}}$ and
$H(r)=\frac{1}{K(r)}$. The nonzero components of Einstein tensor of
cylindrical symmetric interior geometry are
\begin{align}\label{14}
G_{00}&=\frac{H}{K^2 r^2}({K'}r-K),\\\label{15}
G_{11}&=\frac{1}{Hr^2}({H'}r+H),\\\label{16}
G_{22}&=\frac{r}{4H^2K^2}(2{{H''}}HK-H{H'}{K'}-K{H'}^2r-2{K'}H^2+2{H'}KH),
\end{align}
where prime represents the derivative with respect $r$. Now using
Eqs.\eqref{13}-\eqref{16} in Eq.\eqref{11} then we get the following
relations
\begin{align}\label{17}
&\frac{{K'}r-K}{K^2}=-r^2[8\pi\rho-\chi(p-3\rho)],\\\label{18}
&\frac{{H'}r+H}{H K}= -r^2[8\pi p+\chi(3p-\rho)],\\\label{19}
&\frac{r}{4H^2K^2}(2{{H''}}HK-H{H'}{K'}-K{H'}^2r-2{K'}H^2+2{H'}K
H)= -r^2[8\pi p+\chi(3p-\rho)].
\end{align}
Here $G_{33}$=$\alpha^2G_{22}$. The hydrostatic equilibrium equation
follows the nonconservation Eq.\eqref{9} as
\begin{eqnarray}\label{20}
\frac{d
p}{dr}+\frac{{H'}(\rho+p)}{2H}+\frac{\chi}{2(4\pi+\chi)}\left(\frac{dp}{dr}-\frac{d\rho}{dr}\right)=0.
\end{eqnarray}
If $r$ is the gravastar inner radius and $m$ is the mass of the
gravitating source within it, then by using Eq.\eqref{17} we get
\begin{eqnarray}\label{21}
\frac{1}{K}=\frac{8m}{3h}+\chi(\rho-\frac{p}{3})r^2.
\end{eqnarray}
Substituting Eq.\eqref{21} in Eq.\eqref{20}, we obtain
\begin{eqnarray}\label{22}
\frac{d
p}{dr}=-\frac{(\rho+p)}{2}\left[\frac{r^2[-8\pi\rho+\chi(\rho-3p)]}{[\frac{8m}{3h}+\chi(\rho-\frac{p}{3})r^2][1+\frac{\chi}{2(4\pi+\chi)}(1-\frac{d\rho}{d
p})]}\right].
\end{eqnarray}

\section{Interior Geometry}

In this section, we will explore the inner region of gravastar. We
have already discussed that the structure of gravastar is based on
three different regions (a) inner region (b) transitional layer and
(c) outer region. The structure of gravastar is based on the
parameter of the EoS. Also, $r_1=D=r_2$ as the radius of inner and
outer regions with EoS $p=-\rho$ and $p=\rho$, respectively while
$r_1\leq r<r_2$ is the radius of thin shell with EoS $p=0=\rho$. To
find the structural form of the coordinate of metric and their
relationships, we use EoS of the interior region $p=-\rho$. Assume
that density is constant within the interior region so $\rho=Y_0$
\begin{eqnarray}\label{23}
p=-Y_0.
\end{eqnarray}
Now, using Eq.\eqref{23} in Eq.\eqref{17} we obtain
\begin{eqnarray}\label{24}
K=\frac{3}{r^3}\left[\frac{1}{4(2\pi+\chi)Y_0}\right],~~~~~~~
and~~~~~~~K=B H^{-1},
\end{eqnarray}
where $B$ is the integration constant. A relationship between the
spacetime potentials $K$ and $H$ is defined in Eq.\eqref{24}.
Gravitational mass $M(D)$ of the inner region is defined as in given
below
\begin{eqnarray}\label{25}
M(D)=\int^{r_1=D}_04\pi Y_0r^2dr=\frac{4}{3}\pi D^3Y_0.
\end{eqnarray}

\section {Geometry of Thin Shell}

This section provides the shape of shell and we will examine the
influence of pressure on the radius of gravastar. The shell is
composed of ultrarelativistic matter under nonvacuum region. The
state equation for intermediate shell is $p=\rho$. With the help of
$\rho=p$ state equation in Eqs.\eqref{17} and \eqref{18}, one can
get
\begin{eqnarray}\label{26}
\frac{d}{dr}(\ln r)=\frac{2}{r},
\end{eqnarray}
The substitution of EoS $\rho=p$ in Eqs.\eqref{17} and \eqref{19}
provide
\begin{eqnarray}\label{27}
\frac{r}{2}\left(\frac{r{H'}}{2H}+3\right){K'}=r^2.
\end{eqnarray}
The integration of Eq.\eqref{26} yields
\begin{eqnarray}\label{28}
K=r^2+C,
\end{eqnarray}
where $C$ represents integration constant while $r$ is the radius
with range $D\leq r\leq D+\epsilon$, where $\epsilon\ll1$.
Integrating Eq.\eqref{27}, we get
\begin{eqnarray}\label{29}
\frac{{H'}}{H}=-\frac{6}{r}.
\end{eqnarray}
By making use of EoS $p=\rho$ in Eq.\eqref{20}, it follows that
\begin{eqnarray}\label{30}
p=\rho=Fr^6.
\end{eqnarray}
\begin{center}\begin{figure}\centering
\epsfig{file=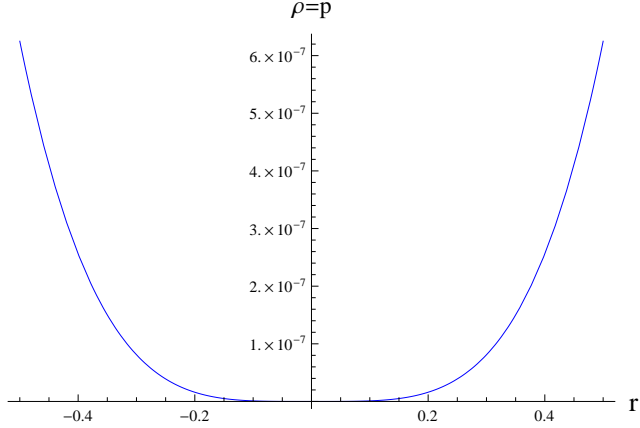,width=0.51\linewidth}\caption{\small{Plot of
$p=\rho$(cm) verses $r$ of the ultrarelativistic fluid in
the shell}}\label{b2}
\end{figure}\end{center}

\section{Junction Conditions}

The configuration of gravastar is based on three different shall,
i.e interior cylindrical shell, thin shell, and exterior cylindrical
shell. In this section, we shall find the condition on an
intermediate thin shell by interior and exterior regions. For smooth
matching, we used Darmois \cite{18} and Israel \cite{17} junction
conditions. Here, we consider the following spacetime
as\begin{eqnarray}\label{31}
ds^2=-\left(1-\frac{r^2}{\beta^2}\right)dt^2+\left(1-\frac{r^2}{\beta^2}\right)^{-1}dr^2+r^2(d\phi^2+\alpha^2dz^2).
\end{eqnarray}
One can observe the continuity of coefficients across the boundary
surface, the existence of their derivative at $r=D$ may not be
continuous. To overcome the effects of discontinuity, we use Israel
formulation. The Lanczos equation \cite{19,20,21,22} is defined as
\begin{eqnarray}\label{32}
S_\iota^\kappa=-\frac{1}{8\pi}(\xi_\iota^\kappa-\delta_\iota^\kappa
\xi_i^i),
\end{eqnarray}
where $S_\iota^\kappa$ expression is used for stress-energy surface
and $\xi_i^i=\Omega_{ij}^+-\Omega_{ij}^-$. Here $(+)$ sign
represents the interior region and $(-)$ sign represents exterior
region. Second fundamental \cite{23,24,25} form of hypersurface is
defined as
\begin{align}\label{34}
\Omega_{ij}^{\pm}&= =-n_\nu^{\pm}\left[\frac{\partial^2y_\nu}{\partial\zeta^i\partial\zeta^j}+\Gamma_{\gamma\delta}^\nu\frac{\partial
y^\gamma}{\partial\zeta^i}\frac{\partial
y^\delta}{\partial\zeta^j}\right]_\Sigma,
\end{align}
where $\zeta^i$ are obtained through the coordinates on the
hypersurface, $y^\gamma$ represents the coordinate of manifold and
$n_\nu^{\pm}$ is the normal over the boundary which is defined as
\begin{eqnarray}\label{35}
n_\nu^{\pm}=\pm\frac{1}{\left|g^{\alpha\beta}\frac{\partial
\ell(r)}{\partial y^\alpha}\frac{\partial \ell(r)}{\partial
y^\beta}\right|^{\frac{1}{2}}}\frac{\partial \ell(r)}{\partial
y^\nu},
\end{eqnarray}
with timelike condition. From Lanczos equation, one can obtain
$S_\iota^\kappa=diag[\vartheta,-\varrho,-\varrho,-\varrho]$. Here
$\vartheta$ denotes the surface density and $\varrho$ denotes the
isotropic pressure over the boundary surface. The values of
$-\varrho$ and $\vartheta$ are
\begin{eqnarray}\label{36}
&&\vartheta=-\frac{1}{4\pi D}[\sqrt{\ell(r)}]_-^+,\\\label{37}
&&\varrho=-\frac{\vartheta}{2}+\frac{1}{16\pi}\left[\frac{\ell'(r)}{\sqrt{\ell(r)}}\right]_-^+.
\end{eqnarray}
Using Eqs.\eqref{36} and \eqref{37} we get
\begin{eqnarray}\label{38}
\vartheta=\frac{1}{4\pi
D}\left[\sqrt{\left(\frac{D^2}{\beta^2}-1\right)}-\sqrt{\frac{4D^2}{3}(2\pi+\chi)\rho_0}~
\right],
\end{eqnarray}
and
\begin{eqnarray}\label{39}
\varrho=\frac{1}{8\pi
D}\left[\frac{\left(\frac{2D^2}{\beta^2}-1\right)}{\sqrt{\left(\frac{D^2}{\beta^2}-1\right)}}
-\frac{\frac{8\rho_0D^2}{3}(2\pi+\chi)}{\sqrt{\frac{4D^2}{3}(2\pi+\chi)\rho_0}}\right].
\end{eqnarray}
The thin shell mass can be calculated via areal density as
\begin{eqnarray}\label{40}
m_s=4\pi
D^2\vartheta=D\left[\sqrt{\frac{4D^2}{3}(2\pi+\chi)\rho_0}-\left(\frac{D^2}{\beta^2}-1\right)~
\right],
\end{eqnarray}
where
\begin{eqnarray}\label{41}
\beta=\sqrt{\frac{D^4}{m_s^2}+\frac{3}{2\rho_0(2\pi+\chi)}+\frac{D^3}{2m_s}\sqrt{\frac{3}{2\rho_0D^2(2\pi+\chi)}}+D^2}.
\end{eqnarray}

\section{Characteristics of Gravastars}

In this segment, we discuss different physical characteristics of
gravastar through graphical representation.

\subsection{Length of Intermediate Shell}

This subsection is devoted to find the thickness of the shell using
interior and exterior radius. We assume that the ratio $\epsilon/D$
is too much less, i.e, $\epsilon/D<<1$. The length of the
intermediate shell can be calculated as \cite{26}
\begin{eqnarray}\label{42}
\ell=\int^{D+\epsilon}_D\sqrt{K}~dr =
\int^{D+\epsilon}_D\sqrt{r^2+C}~dr.
\end{eqnarray}
Integrating the above equation gives us
\begin{eqnarray}\label{43}
\ell=\left[~\frac{1}{2}r\sqrt{r^2+C}+\frac{1}{2}C\ln(r+\sqrt{r^2+C})~\right]^{D+\epsilon}_D.
\end{eqnarray}
\begin{center}\begin{figure}\centering
\epsfig{file= 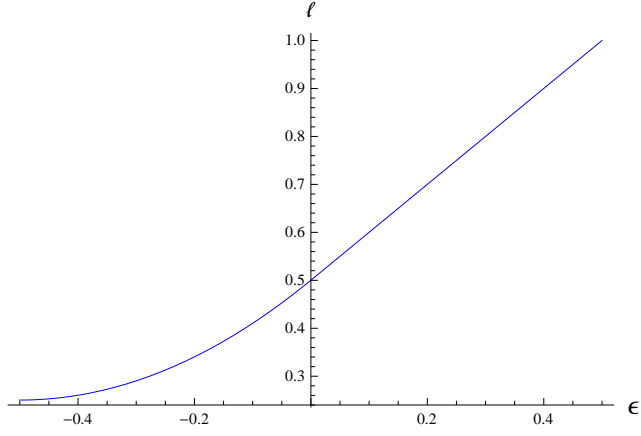,width=0.51\linewidth}\caption{\small{Plot of
 $\ell$(km) verses thickness $\epsilon$ (km) in the gravastar.}}\label{b2}
\end{figure}\end{center}

\subsection{Energy Contents}

\begin{center}\begin{figure}\centering
\epsfig{file= 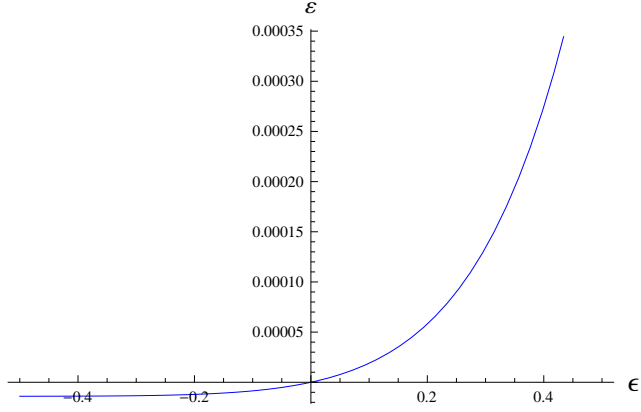,width=0.51\linewidth}\caption{\small{Plot of
energy $\varepsilon$ (J) verses $\epsilon$ (km) in the
gravastar.}}\label{b3}
\end{figure}\end{center}

Energy Contents within the thin shell is
given as \cite{26}
\begin{eqnarray}\label{44}
\varepsilon=\int^{D+\epsilon}_D4\pi\rho r^2~dr=
\int^{D+\epsilon}_D4\pi Hr^8~dr,
\end{eqnarray}
which gives
\begin{eqnarray}\label{45}
\varepsilon=\frac{4}{9}\pi H ~[(D+\epsilon)^9-D^9].
\end{eqnarray}
We can see the behavior of energy contents within the thin shell
against thickness is shown in Fig.\eqref{b3}. Graph shows the
positive and linear relationship.

\subsection{Entropy within the Shell}

By the theory of Mazur and Mottola \cite{1,2}, gravastar has zero disorderness in the interior
cylindrical region. Now we are going to determine the
corresponding entropy \cite{26} by letting $r_1=D$ and $r_2=D+\epsilon$
\begin{eqnarray}\label{46}
S=\int^{D+\epsilon}_D4\pi r^2\aleph(r)\sqrt{K}~dr,
\end{eqnarray}
where $\aleph(r)$ is the entropy density and is defined as
\begin{eqnarray}\label{47}
\aleph(r)=\left(\frac{\alpha
K_B}{h}\right)\left({\frac{P}{2\pi}}\right)^\frac{1}{2},~~~~~~~~~K_B=h=1.
\end{eqnarray}
Using Eqs.\eqref{30} and \eqref{47} in Eq.\eqref{46}, the entropy
within the shell is
\begin{eqnarray}\label{48}
S=(8\pi F)^\frac{1}{2}\alpha\int^{D+\epsilon}_Dr^5\sqrt{r^2+C}~dr,
\end{eqnarray}
which provides
\begin{eqnarray}\label{49}
S=(8\pi
F)^\frac{1}{2}\alpha\left[\frac{1}{105}(r^2+C)^{\frac{3}{2}}(15r^4-12Cr^2+8C^2)\right]^{D+\epsilon}_D.
\end{eqnarray}
\begin{center}\begin{figure}\centering
\epsfig{file= 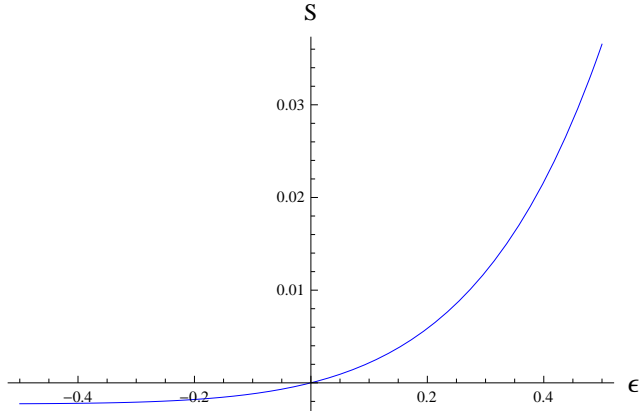,width=0.51\linewidth}\caption{\small{Plot of
the entropy of the shell against the thickness $\epsilon$ (km) of
the shell.}}\label{b4}
\end{figure}
\end{center}

\subsection{State Equation}

The EoS explains the link between density and pressure as
\begin{eqnarray}\label{50}
\varrho=\omega(D)\vartheta,
\end{eqnarray}
substituting the values of $\varrho$ and $\vartheta$ from
Eqs.\eqref{38} and \eqref{39} in above the above equation, it
follows that
\begin{eqnarray}\label{51}
\omega(D)=\frac{\left[\frac{\left(\frac{2D^2}{\beta^2}-1\right)}{\sqrt{\left(\frac{D^2}{\beta^2}-1\right)}}
-\frac{\frac{8\rho_0D^2}{3}(2\pi+\chi)}{\sqrt{\frac{4D^2}{3}(2\pi+\chi)\rho_0}}\right]}{2\left[\sqrt{\left(\frac{D^2}{\beta^2}-1\right)}-\sqrt{\frac{4D^2}{3}(2\pi+\chi)\rho_0}~
\right]},
\end{eqnarray}
we used some approximation $\frac{D^2}{\beta^2}\ll1$ and
$\frac{4}{3}\rho_0(2\pi+\chi)D^2\ll1$ to obtain a real solution,
thus we get
\begin{eqnarray}\label{52}
\omega(D)\approx-1.
\end{eqnarray}

\section{Conclusion}

As a result of the gravitational collapse, the presence of a new
theoretical system may be evident. It can be accomplished by
carefully examining the fundamental concepts of BEC. In BEC the
boson gas is cooled to the point where infinite kinetic energy is
lost by the molecules. We should consider that all boson molecules
are identical with the same quantum spin, and thus do not obey the
exclusive theory of Pauli. Any of the final gravitational collapsing
stages may lead to gravastar, thus suggesting this as an alternative
to the black hole as indicated by \cite{1}.

In this work, we have discussed the geometry of gravastar and their
different characteristics under static cylindrically symmetric
spacetime. The solution of the gravastar is based on these three
cases, first is the interior, second is the thin shell and third is
the exterior region of cylindrically. By matching the first and
third regions, we obtain a mass of thin cylindrical shell. In this
paper, we calculate the gravitational mass of the interior
cylindrical region using EoS $p=-\rho$ under constant density.

We have also discussed the characteristics of gravastar, e.g proper
length of the thin shell, entropy and energy contents within the
shell. In this first hand, we have studied the proper length of a
thin shell. It is found that it depends upon the radius of the
interior and exterior regions. In Fig.\eqref{b2}, we see the
behavior of the proper length of the thin shell against the
thickness. Second is the energy content within the shell,
Fig.\eqref{b3} shows the positive and linear relationship between
energy and thickness. The third that we have discussed is the
entropy, which is the disorderliness in the surface of gravastar. In
Fig.\eqref{b4}, we observe that if thickness increases then entropy
will also increase. Lastly, we calculate the equation of state by
putting some certain state such as $\frac{D^2}{\beta^2}\ll1$ and
$\frac{4}{3}\rho_0(2\pi+\chi)D^2\ll1$.

\vspace{0.3cm}

\section*{Acknowledgments}

ZY and MZB acknowledge the financial support under NRPU from Higher
Education Commission, Pakistan via project No.
8754/Punjab/NRPU/R\&D/HEC/2017. In addition, the work of KB has been supported in part by the JSPS KAKENHI Grant Number JP21K03547.

\end{document}